\documentclass[aps,pre,twocolumn,showpacs]{revtex4}
\usepackage{amsfonts}
\usepackage{amsmath}
\usepackage{amssymb}
\usepackage{graphicx}
\usepackage{color}

\def\kbar{\protect\@kbar}
\def\@kbar{\relax \bgroup
\def\@tempa{\hbox{\raise.73\ht0
\hbox to0pt{\kern.25\wd0\vrule width.5\wd0 height.1pt
depth.1pt\hss}\box0}}\mathchoice{\setbox0\hbox{$\displaystyle
k$}\@tempa}{\setbox0\hbox{$\textstyle
k$}\@tempa}{\setbox0\hbox{$\scriptstyle
k$}\@tempa}{\setbox0\hbox{$\scriptscriptstyle k$}\@tempa}\egroup}

\begin{document}

\title{\textbf{Kicked Hall Systems: Quantum-Dynamical and Spectral Manifestations of Generic Superweak Chaos}}
\author{Itzhack Dana and Kazuhiro Kubo}
\affiliation{Minerva Center and Department of Physics, Bar-Ilan University, Ramat-Gan 52900, Israel}

\begin{abstract}
Classical ``kicked Hall systems"(KHSs), i.e., periodically kicked charges in the presence of uniform magnetic and electric fields that are perpendicular to each other and to the kicking direction, have been introduced and studied recently. It was shown that KHSs exhibit, under generic conditions, the phenomenon of ``superweak chaos"(SWC), i.e., for small kick strength $\kappa$ a KHS behaves as if this strength were effectively $\kappa^2$ rather than $\kappa$. Here we investigate quantum-dynamical and spectral manifestations of this generic SWC. We first derive general expressions for quantum effective Hamiltonians for the KHSs. We then show that the phenomenon of quantum antiresonance (QAR), i.e., ``frozen" quantum dynamics with flat quasienergy (QE) bands, takes place for integer values of a scaled Planck constant $\hbar_{\rm s}$ and under the same generic conditions for SWC. This appears to be the most generic occurrence of QAR in quantum systems. The vicinity of QAR is shown to correspond semiclassically to SWC. A global spectral manifestation of SWC is the fact that a scaled QE spectrum as function of $\hbar_{\rm s}$, at fixed small value of $\kappa /\hbar_{\rm s}$, features an approximately ``doubled" structure. In the case of standard (cosine) potentials, this structure is that of a universal (parameters-independent) double Hofstadter butterfly. Also, for standard potentials and for small $\hbar_{\rm s}$ (semiclassical regime), the evolution of the kinetic-energy expectation value exhibits a relatively slow quantum-diffusive behavior having universal features. These approximate spectral and quantum-dynamical universalities agree with predictions from the effective Hamiltonian.
\end{abstract}


\maketitle

\begin{center}
\textbf{I. INTRODUCTION}
\end{center}

Low-dimensional quantum systems, whose classical counterparts are nonintegrable with chaotic dynamics, exhibit a rich variety of behaviors both in semiclassical and fully quantum regimes. Two basically different classes of such systems have become paradigmatic in the field of ``Quantum Chaos". The first class includes the kicked rotor (KR) and several variants of it \cite{qc,bc,jmg,mmp,df,dr,qkr,fmi,qar,fgr,dd,dvr,dle,qre,qrre}. The classical KR features unbounded chaotic diffusion on a cylindrical phase space for sufficiently strong nonintegrability or kicking parameter \cite{bc,jmg,mmp,df}. Quantally, this diffusion is suppressed due to dynamical localization in angular-momentum space for generic (irrational) values of a scaled (dimensionless) Planck constant $\hbar_{\rm s}$ \cite{qc,qkr,qar,qre}. The dynamical localization is a consequence of an essentially discrete quasienergy (QE) spectrum \cite{qkr}. For rational $\hbar_{\rm s}$, the QE spectrum is band continuous, leading to the diametrically opposite phenomenon of quantum resonance (QR) \cite{fmi,fgr,dd,dvr}, i.e., a quadratic growth in time of the expectation value of the quantum KR kinetic energy. Experimental realizations of both dynamical localization \cite{dle} and QR phenomena \cite{qre,qrre} were achieved using atom-optics methods with cold atoms or Bose-Einstein condensates.

A second class of systems, introduced by Zaslavsky and coworkers \cite{wm,lw} and subsequently generalized in other works \cite{da,dk,dh,d,prk}, are charged particles periodically driven or kicked by a spatially periodic potential in a direction perpendicular to a uniform magnetic field ${\bf B}$. Unlike the classical KR, these systems can exhibit, at least for some parameter values, an unbounded chaotic diffusion on an infinite  ``stochastic web" in the phase plane for arbitrarily small nonintegrability strength \cite{wm,lw,d,prk}. Quantally, the QE spectrum of such weak-chaos web systems was numerically shown to have a fractal structure for generic $\hbar_{\rm s}$ \cite{fqd}, leading to quantum diffusion \cite{fqd,wa,dd1}, i.e., an almost linear increase in time of the kinetic-energy expectation value. QR again occurs in these systems for rational $\hbar_{\rm s}$.

Recently, another class of systems has been introduced and their classical dynamics was studied. These are the ``kicked Hall systems" (KHSs) \cite{bhd}, obtained from the Zaslavsky systems by adding a uniform electric field ${\bf E}$ perpendicular to both the magnetic field ${\bf B}$ and the kicking direction. It was shown \cite{bhd} that for resonant values of $B$ and $E$ and for small kicking strength $\kappa$ there exists a \emph{generic} family of periodic kicking potentials for which the Hall effect from ${\bf B}$ and ${\bf E}$ significantly suppresses the weak chaos in the Zaslavsky systems, replacing it by \emph{``superweak"} chaos (SWC). This means that the system behaves as if the kicking strength were $\kappa ^2$ rather than $\kappa$. Classical manifestations of SWC are a decrease in the instability of periodic orbits, a narrowing of the chaotic layers, and slower chaotic diffusion on stochastic webs, relative to the ordinary weak-chaos case ($E=0$) \cite{bhd}.

In this paper, we investigate quantum-dynamical and spectral manifestations of SWC in KHSs by restricting ourselves, for simplicity, to stochastic webs with square rotational symmetry. The content and organization of the paper are as follows. In Sec. II, we present a summary of relevant properties of classical KHSs (see more details in Ref. \cite{bhd}). In Sec. III, the basic evolution operator for the quantum KHS is given and general expressions for quantum effective Hamiltonians are derived. In Sec. IV, we show that for integer values of $\hbar_{\rm s}$ and for the same generic family of kicking potentials for which SWC occurs, there takes place the phenomenon of quantum antiresonance (QAR), i.e., frozen quantum dynamics with flat (infinitely degenerate) QE bands. This QAR in KHSs is much more generic than the rare QAR occurring in the KR \cite{fmi} (or variants of it \cite{qar}) and in the Zaslavsky systems \cite{d,dd1}, as a special singular case of QR. We also show that the vicinity of QAR ($\hbar_{\rm s}$ close to integers) corresponds semiclassically to SWC. In Sec. V, we show the following global spectral manifestation of SWC: A scaled QE spectrum as function of $\hbar_{\rm s}$, at fixed small value of $\kappa/\hbar_{\rm s}$, features an approximately ``doubled" structure. In the case of standard (cosine) potentials, this structure is that of a universal (parameters-independent) double Hofstadter butterfly. In Sec. VI, we study numerically the evolution of the kinetic-energy expectation value of the KHS for standard potentials in a semiclassical regime (small irrational $\hbar_{\rm s}$). This evolution is found to exhibit an approximate quantum-diffusive behavior which is slower than that in the ordinary weak-chaos case and has universal features under variations of the electric field. The observed approximate universalities are in accordance with predictions from the effective Hamiltonian. A summary and conclusions are presented in Sec. VII.

\begin{center}
\textbf{II. SUMMARY OF PROPERTIES OF CLASSICAL KHSs}
\end{center}

The KHS is a charged particle in uniform magnetic and electric fields, ${\bf B}=B{\bf \hat{z}}$ and ${\bf E}=E{\bf \hat{y}}$ respectively, and periodically kicked by a spatially periodic potential $V(x)$ in the $x$ direction. Assuming, without loss of generality, a particle of unit mass and charge, the Hamiltonian is:
\begin{equation}
H=\frac{\mathbf{\Pi }}{2}^{2}-Ey+\kappa V(x)\sum_{s=-\infty }^{\infty }\delta
(t-sT),\ \ \   \label{H}
\end{equation}
where $\mathbf{\Pi}= \mathbf{p}-\mathbf{B}\times\mathbf{r}/(2c)$ is the
kinetic momentum, $\kappa$ is a nonintegrability parameter, and $T$ is the time period. It is useful to express (\ref{H}) in the two natural degrees of freedom
in a magnetic field \cite{jl}, given by the independent conjugate pairs
$(x_{\mathrm{c}},y_{\mathrm{c}})$ (coordinates of the cyclotron-orbit center)
and $(u=\Pi _{x}/\omega ,v=\Pi _{y}/\omega )$, with $\omega =B/c$ being the cyclotron angular velocity. From simple geometry one has $x=x_{\mathrm{c}}-v$ and $y=y_{\mathrm{c}}+u$. Defining the variable $u^{\prime }=u-E/\omega ^{2}$, which we re-denote by $u$, the Hamiltonian (\ref{H}) can then be expressed as
follows: 
\begin{equation}
H=\omega ^{2}(u^{2}+v^{2})/2-Ey_{\mathrm{c}}+\kappa V(x_{\mathrm{c}
}-v)\sum_{s=-\infty }^{\infty }\delta (t-sT),  \label{eKHO} 
\end{equation}
where a constant $E^{2}/(2\omega ^{2})$ was omitted. Choosing units such that $\omega =1$ from now on, the conjugate pairs above have Poisson brackets $\{y_{
\mathrm{c}},x_{\mathrm{c}}\}=\{u,v\}=1$. From the Hamilton
equation $\dot{x}_{\mathrm{c}}=-\partial
H/\partial y_{\mathrm{c}}=E$, we see that $x_{\mathrm{c}}$ evolves
linearly in time (Hall effect): 
\begin{equation}\label{xct}
x_{\mathrm{c}}=x_{\mathrm{c}}^{(0)}+Et, 
\end{equation}
Using Eq. (\ref{xct}), we see that the Hamiltonian (\ref{eKHO}) is just that of a harmonic oscillator [in the conjugate pair $(u,v)$] periodically kicked by a time modulated potential $V(x_{\mathrm{c}}^{(0)}+Et-v)$.
For $E=0$, $x_{\mathrm{c}}$ is a constant of the motion.

From $\{ u,v\} =1$, the Hamilton equations for $(u,v)$ are $\dot{u}
=\partial H/\partial v$ and $\dot{v}=-\partial
H/\partial u$, where $H$ is given by (\ref{eKHO}) with (\ref{xct}).
Integrating the latter equations from $t=sT-0$ to $t=(s+1)T-0$ and denoting $u_s=u(t=sT-0)$, $v_s=v(t=sT-0)$, one obtains the one-period Poincar\'{e} map for the KHS: 
\begin{equation}  \label{cMh}
M_{\gamma ,\eta}:\ z_{s+1}=[z_s+\kappa f(x_{\mathrm{c}}^{(0)}+s
\eta-v_s)]e^{-i\gamma},
\end{equation}
where $z_s=u_s+iv_s$, $f(x)=-dV/dx$, $\gamma =\omega T=T$, and $\eta =ET$. We assume the period of $V(x)$ to be $2\pi$, without loss of generality, and that $\gamma$ and $\eta$ satisfy the resonance conditions: 
\begin{equation}  \label{rge}
\frac{\gamma}{2\pi}=\frac{m}{n},\ \ \ \ \frac{\eta}{2\pi}=\frac{k}{\ell},
\end{equation}
where $(m,n)$ and $(k,\ell )$ are two pairs of coprime integers. Let $r=\mathrm{lcm}(n,\ell )$ be the least common multiple of $n$ and $\ell$. Then, the map from $z_s$ to $z_{s+r}$ is given by
\begin{equation}  \label{cMhb}
M_{\gamma ,\eta}^r:\ z_{s+r}=z_s+\kappa\sum_{j=0}^{r-1}f(x_{\mathrm{c}
}^{(0)}+(s+j)\eta-v_{s+j})e^{ij\gamma}
\end{equation}
and has a translationally invariant form under $s\rightarrow s\pm r$. Also, the map (\ref{cMhb}) is the smallest iterate of the map (\ref{cMh}) that is a near identity ($z_{s+r}\approx z_s$) for small $\kappa$. Thus, (\ref{cMhb}) may be considered as the basic map for the system. 

The map (\ref{cMh}) for $n=1,2$ ($\gamma =0,\pi$) is integrable for all $\eta$, so that chaos may emerge only for $n>2$. We say that the map (\ref{cMhb}) for $n>2$ and small $\kappa\ll 1$ exhibits SWC if its expansion in powers of $\kappa$ starts from  $\kappa ^2$, 
\begin{equation}\label{SWC}
M_{\gamma ,\eta ,r}:\ z_{s+r}=z_s+O(\kappa ^2).
\end{equation}
This is unlike ordinary weak chaos, with $z_{s+r}=z_s+O(\kappa )$.

Given  the general family of $2\pi$-periodic potentials with finite Fourier
expansion, 
\begin{equation}\label{V}
V(x)=\sum_{g=-N}^NV_g\exp{(igx)},\ \ V_0=0,
\end{equation}
one can show the following \cite{bhd}. For $E=0$, Eq. (\ref{SWC}) holds only if $n$ is even and the function $V(x_{\mathrm{c}}-v)$ is odd: $V(x_{\mathrm{c}}+v)=-V(x_{\mathrm{c}}-v)$. For $E\neq 0$, with the resonance conditions (\ref{rge}), let us write $n/\ell =n^{\prime}/\ell ^{\prime}$, where $(n^{\prime},\ell ^{\prime})$ are coprime integers. Then, if 
\begin{equation}  \label{cE}
\ell ^{\prime}> N,
\end{equation}
Eq. (\ref{SWC}) for SWC holds for \emph{arbitrary} potential (\ref{V}) and initial value $x_{\mathrm{c}}^{(0)}$ in Eq. (\ref{cMhb}), independently of the parity of $n$. Thus, unlike the case of $E=0$, SWC for $E\neq 0$ occurs under quite generic conditions.

From now on, we shall restrict ourselves to the case of $\gamma =\pi/2$, i.e., $m/n=1/4$ in Eq. (\ref{rge}), corresponding to chaotic motion on stochastic webs having translational invariance in both $u$ and $v$ with period $2\pi$ and an approximately square web cell for small $\kappa$; see, e.g., Fig. 1. This case was extensively considered in the study of classical KHSs \cite{bhd}.

\begin{figure}[tbp]
\includegraphics[width=8.5cm,trim=0.5cm 2cm 0cm 2cm]{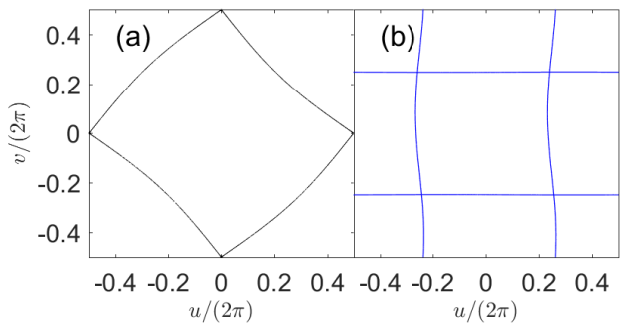}
\caption{(Color online) Portions of stochastic webs in the $2\pi\times 2\pi$ unit cell of periodicity $-\pi\leq u,v <\pi$ for $V(x)=-\cos (x)$, $\gamma =\pi /2$, $x_{\rm c}^{(0)}=0$, and: (a) $\kappa =0.6$, $\eta=0$ (ordinary weak-chaos case); (b) $\kappa =0.1$, $\eta =4\pi /3$ (SWC case). Notice the approximately square web cell in both cases. The web-cell area in case (a) is approximately twice that in case (b).}
\label{fig1}
\end{figure}

\begin{center}
\textbf{III. QUANTUM EFFECTIVE HAMILTONIANS FOR KHSs}
\end{center}

\begin{center}
\textbf{A. KHS basic evolution operator}
\end{center}

The quantum analogs of the Poisson brackets $\{y_{
\mathrm{c}},x_{\mathrm{c}}\}=\{u,v\}=1$ are the commutators $[\hat{y}_{
\mathrm{c}},\hat{x}_{\mathrm{c}}]=[\hat{u},\hat{v}]=i\hbar$, where the quantum variables (operators) are indicated by carets as usual. From Eq. (\ref{eKHO}) (with $\omega =1$ in our units), the quantum Hamiltonian is written as
\begin{equation}\label{qH}
\hat{H}=\hbar (\hat{a}^{\dagger}\hat{a}+1/2)-E\hat{y}_{\mathrm{c}}+\kappa V(\hat{x}_{\mathrm{c}
}-\hat{v})\sum_{s=-\infty }^{\infty }\delta (t-sT),  
\end{equation} 
where $\hat{a}=(\hat{v}-i\hat{u})/\sqrt{2\hbar}$. The one-period evolution operator for the Hamiltonian (\ref{qH}), from $t=sT-0$ to $t=(s+1)T-0$, is given by
\begin{equation}\label{U}
\hat{U}=\hat{U}_{\gamma}\hat{U}'_{\eta}\exp\left[-i\mu V(\hat{x}_{\mathrm{c}
}-\hat{v}) \right],
\end{equation}
where $\hat{U}_{\gamma}=\exp\left[-i\gamma(\hat{a}^{\dagger}a+1/2) \right]$ ($\gamma =\omega T=T$), $\hat{U}'_{\eta}=\exp(i\eta \hat{y}_{\mathrm{c}}/\hbar )$ ($\eta =ET$), and $\mu =\kappa /\hbar$. The operator $\hat{U}_{\gamma}$ is a rotation by angle $\gamma$ in the $(u,v)$ phase plane: $\hat{U}_{\gamma}f(\hat{a}^{\dagger},\hat{a})\hat{U}^{-1}_{\gamma}=f(\hat{a}^{\dagger}e^{-i\gamma},\hat{a}e^{i\gamma})$, for arbitrary function $f(\hat{a}^{\dagger},\hat{a})$ \cite{amp}. In the case assumed in this paper, i.e., $\gamma =\pi /2$, this is a clockwise rotation by $\pi /2$: $v\rightarrow u\rightarrow -v\rightarrow -u\rightarrow v$. From $[\hat{y}_{\mathrm{c}},\hat{x}_{\mathrm{c}}]=i\hbar$, one has $\hat{y}_{\mathrm{c}}=-i\hbar d/d{x}_{\mathrm{c}}$, so that $\hat{U}'_{\eta}=\exp(-\eta d/d{x}_{\mathrm{c}})$, a translation of $\hat{x}_{\mathrm{c}}$ by $-\eta$. Using all this and Eq. (\ref{rge}), one can then write the basic evolution operator for the KHS, $\hat{U}^r$ [$r=\mathrm{lcm}(n=4,\ell )$, see Sec. II], corresponding to the basic classical map (\ref{cMhb}):
\begin{equation}\label{Urf}
\hat{U}^r = \hat{U}_r (-1)^{\ell '}\exp (-2\pi n'k d/d{x}_{\mathrm{c}}) ,
\end{equation}
\begin{equation}\label{Ur}
\hat{U}_r=\prod_{j=1}^r \exp\left[-i\mu V\left(\hat{x}_{\mathrm{c}}-j\eta
-\hat{v}_j\right)\right] , 
\end{equation}
where the factors in the product are arranged from left to right in order of increasing $j$ and we defined $\hat{v}_1=\hat{u}$, $\hat{v}_2=-\hat{v}$, $\hat{v}_3=-\hat{u}$, $\hat{v}_4=\hat{v}$, with $\hat{v}_j$ being periodic in $j$ with period $4$. The two terms besides $\hat{U}_r$ in Eq. (\ref{Urf}) are as follows: $(-1)^{\ell '}=\hat{U}_{\gamma}^r$, since $\hat{U}_{\gamma}^4=-1$ and $r=4\ell '$; $\exp (-2\pi n'k d/d{x}_{\mathrm{c}})=(\hat{U}'_{\eta})^r$ from Eq. (\ref{rge}) and $r=n'\ell$. Since the first term is just a constant and we shall consider only wave functions in the $(u,v)$ degree of freedom, independent of $(x_{\mathrm{c}},y_{\mathrm{c}})$, these terms will be ignored from now on, so that $\hat{U}^r$ is given by just Eq. (\ref{Ur}) with $\hat{x}_{\mathrm{c}}$ replaced by a constant $x_{\mathrm{c}}$.
   
\begin{center}
\textbf{B. General quantum effective Hamiltonians}
\end{center}

The basic evolution operator, given Eq. (\ref{Ur}), is a unitary one and can then be formally written as
\begin{equation}\label{Ure}
\hat{U}_r = \exp (-i\mu \hat{H}_{\rm e}),
\end{equation}
where $\hat{H}_{\rm e}$ is a Hermitian operator, the quantum effective Hamiltonian. We shall now derive an expression for $\hat{H}_{\rm e}$. Assuming a general potential (\ref{V}), the argument of the exponent under the product sign in Eq. (\ref{Ur}) is given by
\begin{eqnarray}\label{Vt}
& & \hat{O}_j =  -i\mu V(x_{\mathrm{c}}-j\eta
-\hat{v}_j) \notag \\  & = & -i\mu  \sum_{g=-N}^N V_g\exp [ig(x_{\mathrm{c}}-j\eta )] \exp(-ig\hat{v}_j).
\end{eqnarray}
Now, given two operators $\hat{A}$ and $\hat{B}$, one has \cite{rmw}
\begin{eqnarray}\label{id}
& & \exp(\hat{A}) \exp(\hat{B}) = \exp\left(\hat{A} + \hat{B}+\frac{1}{2}[\hat{A},\hat{B}] \right. \notag \\
& & \left. +\frac{1}{12}[\hat{A},[\hat{A},\hat{B}]]+
\frac{1}{12}[[\hat{A},\hat{B}],\hat{B}]+\dots \right) , 
\end{eqnarray} 
involving a series of repeated commutators on the right-hand side. Equation (\ref{id}) can be applied to derive systematically an expansion for $\hat{H}_{\rm e}$ in Eq. (\ref{Ure}) as follows. From the definition above of $\hat{v}_j$, we see that if $\hat{v}_j=\hat{u}$ or $\hat{v}_j=\hat{v}$ (up to sign) then $\hat{v}_{j+1}=\hat{v}$ or $\hat{v}_{j+1}=\hat{u}$ (up to sign), respectively. Therefore, the commutator $[\hat{O}_j,\hat{O}_{j+1}]$ of two adjacent operators (\ref{Vt}) will be a linear combination of commutators of the form
\begin{equation}\label{c1}
\left[i\mu e^{ig_1\hat{u}},i\mu e^{ig_2\hat{v}}\right]=2i\mu^2\sin (g_1g_2\pi\hbar_{\rm s}) e^{i(g_1\hat{u}+g_2\hat{v})}
\end{equation}        
for integers $g_1$ and $g_2$, after using Eq. (\ref{id}) with $[\hat{u},\hat{v}]=i\hbar$ and denoting $\hbar_{\rm s}=\hbar/(2\pi )$. More generally, for integers $g_1$, $g_2$, $g_3$, $g_4$,
\begin{eqnarray}\label{c2}
& & \left[i\mu e^{i(g_1\hat{u}+g_2\hat{v})},i\mu e^{i(g_3\hat{u}+g_4\hat{v})}\right] \\ & & = 2i\mu^2\sin [(g_1g_4-g_2g_3)\pi\hbar_{\rm s}] e^{i(g_1+g_3)\hat{u}+i(g_2+g_4)\hat{v}}. \notag
\end{eqnarray}
We also note that for non-zero integer $a$ one has
\begin{equation}\label{id1}
\sin (a\pi\hbar_{\rm s})=J(a;\hbar_{\rm s})\sin (\pi\hbar_{\rm s}),
\end{equation}
where the function $J(a;\hbar_{\rm s})$ does not vanish for integer $\hbar_{\rm s}$. It is then easy to see from Eqs. (\ref{c1})-(\ref{id1}) that the repeated commutators of operators (\ref{Vt}) in Eq. (\ref{id}) imply the following general expansion for the quantum effective Hamiltonian:
\begin{equation}\label{Hee}
\hat{H}_{\rm e}=\hat{H}_0(\hat{u},\hat{v})+\sum_{\imath =1}^{\infty}\epsilon ^{\imath}\hat{H}_{\imath }(\hat{u},\hat{v};\hbar_{\rm s}),
\end{equation}
where
\begin{equation}\label{H0}
\hat{H}_0(\hat{u},\hat{v})=\sum_{j=1}^r V(x_{\mathrm{c}}-j\eta
-\hat{v}_j),
\end{equation}
\begin{equation}\label{eps}
\epsilon =\mu \sin (\pi\hbar_{\rm s}) = \frac{\kappa}{2}\frac{\sin (\pi\hbar_{\rm s})}{\pi\hbar_{\rm s}},
\end{equation}
and $\hat{H}_{\imath}(\hat{u},\hat{v};\hbar_{\rm s})$ is $2\pi$-periodic in both $(\hat{u},\hat{v})$ and non-vanishing for integer $\hbar_{\rm s}$. The $\imath$th term in the expansion (\ref{Hee}) is a linear combination of (repeated) commutators, each involving $\imath +1$ operators (\ref{Vt}) [for example, $[\hat{A},[\hat{A},\hat{B}]]$ in Eq. (\ref{id}) involves three operators]. An explicit general expression for $\hat{H}_1(\hat{u},\hat{v};\hbar_{\rm s})$ is derived in Appendix A, see Eq. (\ref{h1_generic}) there.   

\begin{center}
\textbf{IV. QAR IN KHSs}
\end{center}

\begin{center}
\textbf{A. QAR}
\end{center}

The QAR phenomenon in time-periodic systems occurs when the one-period evolution operator is identically equal to a constant phase factor for some parameter values \cite{fmi,qar,d,dd1}. This implies a frozen quantum dynamics, i.e., no wave packet evolves in time. In the case of the operator (\ref{Ure}), equal to a phase factor, this factor can be chosen to be $1$ without loss of generality. QAR then occurs only if $\hat{H}_{\rm e}$ vanishes identically. From Eq. (\ref{Hee}), we see that $\hat{H}_{\rm e}=0$ provided two conditions are satisfied: 1) $\epsilon =0$, i.e., $\hbar_{\rm s}$ is integer from Eq. (\ref{eps}); 2) $\hat{H}_0(\hat{u},\hat{v})=0$. Using Eq. (\ref{H0}) with Eq. (\ref{V}) and $r=4\ell '$, we get
\begin{eqnarray}\label{H0e}
& & \hat{H}_0(\hat{u},\hat{v}) = \sum_{j=1}^r\sum_{g=-N}^N  V_g e^ {ig(x_{\mathrm{c}}-j\eta )} e^{-ig\hat{v}_j} \notag \\
& = & \sum_{g=-N}^N  V_g e^{igx_{\mathrm{c}}}\sum_{\bar{n}=1}^4\sum_{l=0}^{\ell '-1}e^{-ig[(4l+\bar{n})\eta +\hat{v}_{4l+\bar{n}}]} \\ 
& = & \sum_{g=-N}^N  V_g e^{igx_{\mathrm{c}}}\sum_{\bar{n}=1}^4 e^{-ig(\bar{n}\eta +\hat{v}_{\bar{n}})}\frac{1-e^{-2\pi ikn'g}}{1-e^{-2\pi ikn'g/\ell '}}, \notag
\end{eqnarray}
where we used the periodicity of $\hat{v}_{j}$ with period $n=4$ and the fact that $4\eta=2\pi kn/\ell =2\pi kn'/\ell '$ [see Sec. II, in particular Eq. (\ref{rge})] to perform the sum over $l$. This is a geometric sum, equal to the ratio in the last line of Eq. (\ref{H0e}). Clearly, this ratio is identically zero for all $g$ only if $\ell '>N$. The latter is precisely the SWC condition (\ref{cE}). We thus see that QAR in KHSs occurs for integer $\hbar_{\rm s}$ and under the same generic conditions as SWC, i.e., for general potential (\ref{V}) with $N<\ell '$ and for arbitrary constant $x_{\mathrm{c}}$. 

\begin{center}
\textbf{B. QAR vicinity and SWC}
\end{center}

Consider the close vicinity of QAR, i.e., $\hbar_{\rm s}$ close to an integer value $\hbar_{\rm s}^{(0)}$ (assumed to be odd, for simplicity and without loss of generality): $\hbar_{\rm s}=\hbar_{\rm s}^{(0)}-\delta$, $0<\delta\ll 1$.
Then, from Eq. (\ref{eps}),
\begin{equation}\label{deps}
\epsilon\approx \kappa '=\frac{\kappa\delta}{2(\hbar_{\rm s}^{(0)}-\delta )}.  
\end{equation}
Since $\hat{H}_0(\hat{u},\hat{v})=0$ under the QAR condition $\ell '>N$ (see Sec. IVA), the evolution operator (\ref{Ure}), with $\mu =\kappa /\hbar$ and the expansion (\ref{Hee}), is approximately given by
\begin{equation}\label{Ured}
\hat{U}_r \approx\exp \left[-i\frac{(\kappa ')^2}{\hbar '}\hat{H}_1(\hat{u},\hat{v};\hbar_{\rm s})\right],\ \hbar '=\frac{\hbar\delta}{2(\hbar_{\rm s}^{(0)}-\delta )},
\end{equation}
after using Eq. (\ref{deps}). Equation (\ref{Ured}) is a quantum map deviating from the identity by quantities of order $(\kappa ')^2$, $\kappa '\ll 1$. This quantum map is precisely a semiclassical approximation of the classical SWC map (\ref{SWC}) with $\kappa =\kappa '$ and effective small Planck constant $\hbar '\ll 1$. Thus, the QAR vicinity corresponds to semiclassical SWC.   

\begin{center}
\textbf{V. QE SPECTRA}
\end{center}

\begin{center}
\textbf{A. QE eigenvalue problem}
\end{center}

The eigenvalue problem for the basic evolution operator (\ref{Ur}) is:
$\hat{U}_r|\Psi_{\cal E}\rangle =\exp(-i{\cal E} ) |\Psi_{\cal E}\rangle$, where the phase ${\cal E}$ is the QE determining the eigenvalues $\exp(-i{\cal E} )$. To study the QE spectra, we use the formalism in Ref. \cite{d1}, which we briefly summarize here. Clearly, $\hat{U}_r$ in Eq. (\ref{Ur}) commutes with translations by $2\pi$ in $\hat{u}$ and $\hat{v}$. Since $[\hat{u},\hat{v}]=2\pi i\hbar_{\rm s}$, one has $\hat{u}=2\pi i\hbar_{\rm s}d/dv$ and $\hat{v}=-2\pi i\hbar_{\rm s}d/du$, so that these translations are given by the operators $\hat{D}_0 =\exp(i\hat{v}/\hbar_{\rm s})$ and $\hat{D}_1 =\exp(-i\hat{u}/\hbar_{\rm s})$. In general, the latter operators do not commute. However, for rational $\hbar_{\rm s}=q/p$, where $q$ and $p$ are coprime integers, the operators $\hat{D}_1$ and $\hat{D}_2=\hat{D}_0^q=\exp(ip\hat{v})$ commute and, of course, they commute also with $\hat{U}_r$. The simultaneous QE eigenstates of $\hat{U}_r$, 
$\hat{D}_1$, and $\hat{D}_2$ in the $v$-representation can be written as \cite{d1}:
\begin{eqnarray}\label{qes}
\langle v|\Psi_{b,\mathbf{w}}\rangle & = &\sum_{d=0}^{p-1}\phi_b(d;\mathbf{w})\sum_{l=-\infty}^{\infty} e^{il(w_1/q+2\pi d/p)} \notag \\
& \times & \delta (v-w_2+2\pi l/p) . 
\end{eqnarray} 
Here the index $b=1,\dots ,p$ labels $p$ QE bands ${\cal E}_b(\mathbf{w})$, where $\mathbf{w}=(w_1,w_2)$ is a Bloch wave vector ranging in the Brillouin zone $0\leq w_1<2\pi q/p$, $0\leq w_2<2\pi /p$; $\left\{ \phi_b(d;\mathbf{w})\right\}_{d=0}^{p-1}$, $b=1,\dots ,p$, are $p$ independent vectors of coefficients. Assuming the QE eigenvalues to be all different at any fixed $\mathbf{w}$, i.e., $\exp[-i{\cal E}_b(\mathbf{w})]\neq \exp[-i{\cal E}_{b'}(\mathbf{w})]$ for $b\neq b'$, it is easy to see that each QE band ${\cal E}_b(\mathbf{w})$ is $q$-fold degenerate. In fact, the $q$ operators $\hat{D}_0^{\jmath}$, $\jmath =0,...,q-1$, commute with $\hat{U}_r$ but not with $\hat{D}_1$. Thus, the $q$ states $\hat{D}_0^{\jmath}|\Psi_{b,\mathbf{w}}\rangle$ are all different and are degenerate eigenstates belonging to QE band $b$.

The eigenvalue problem for $\hat{U}_r$ can be written as that of a $p\times p$ unitary matrix in the basis of general states (\ref{qes}), as follows. We first define the operator $\hat{{\mathcal U}}_r=\hat{S} \hat{U}_r \hat{S}^{\dagger}$, where $\hat{S}=\exp \left[-i\mu V\left(x_{\mathrm{c}}-\hat{v}\right)\right]$. Then, using Eq. (\ref{Ur}), we find that
\begin{eqnarray}\label{cUr}
\hat{{\mathcal U}}_r &=& 
\prod_{j=0}^{r-1} \exp\left [-i\mu V\left(x_{\mathrm{c}}-j\eta-\hat{v}_j\right)\right] \notag \\
&=& \prod_{j=0}^{r/2-1} \hat{U}^{(j)}_{\rm KH}(\hat{u},\hat{v}) ,
\end{eqnarray}
where
\begin{eqnarray}\label{UKH}
\hat{U}^{(j)}_{\rm KH}(\hat{u},\hat{v}) &=& \exp\left[-i\mu V\left(x_{\mathrm{c}}-2j\eta-(-1)^j\hat{v}\right)\right] \\
& \times & \exp\left[-i\mu V\left(x_{\mathrm{c}}-(2j+1)\eta-(-1)^j\hat{u}\right)\right] \notag
\end{eqnarray}          
is a generalized ``kicked Harper" evolution operator \cite{d1}. In the basis (\ref{qes}), the operator (\ref{UKH}) is represented by a $p\times p$ $\mathbf{w}$-dependent unitary matrix ${\mathbf M}^{(j)}_{\rm KH}({\mathbf w})$, whose elements are given in Ref. \cite{d1} and, in a more explicit and compact form, in Appendix B. Then, the operator (\ref{cUr}) is represented by the $p\times p$ unitary matrix
\begin{equation}\label{McUr}
{\mathbf M}_r({\mathbf w})=\prod_{j=0}^{r/2-1}{\mathbf M}^{(j)}_{\rm KH}({\mathbf w}),
\end{equation}
whose diagonalization gives the QE eigenvalues of $\hat{{\mathcal U}}_r$. By definition of $\hat{{\mathcal U}}_r$, its eigenvalues are the same as those of $\hat{U}_r$. As explained in Appendix B, the eigenstates $\langle v|\bar{\Psi}_{b,\mathbf{w}}\rangle$ of $\hat{{\mathcal U}}_r$ are given by Eq. (\ref{qes}) with $\phi_b(d;\mathbf{w})$ replaced by $\exp (-idw_2)\bar{\phi}_b(d;\mathbf{w})$, where $\left\{ \bar{\phi}_b(d;\mathbf{w})\right\}_{d=0}^{p-1}$ are the eigenvectors of the matrix (\ref{McUr}). Finally, the eigenstates of $\hat{U}_r$ are obtained as $|\Psi_{b,\mathbf{w}}\rangle =\hat{S}^{\dagger}|\bar{\Psi}_{b,\mathbf{w}}\rangle$.   

\begin{center}
\textbf{B. SWC and nearly doubled global spectrum}
\end{center}

Under the classical SWC condition (\ref{cE}), the leading term (\ref{H0}) in the expansion (\ref{Hee}) vanishes (see Sec. IVA), so that for sufficiently small $\mu =\kappa /\hbar$ the basic operator (\ref{Ure}) is approximately given by
\begin{equation}\label{Ures}
\hat{U}_r \approx\exp \left[-i\mu \epsilon \hat{H}_1(\hat{u},\hat{v};\hbar_{\rm s})\right],
\end{equation}
where an expression for $\hat{H}_1(\hat{u},\hat{v};\hbar_{\rm s})$ is derived in Appendix A, see Eq. (\ref{h1_generic}). Let us now assume that $\ell ' >2N$, a condition stronger than the SWC one (\ref{cE}). Then, the second sum in Eq. (\ref{h1_generic}) does not appear, so that $\hat{H}_1(\hat{u},\hat{v};\hbar_{\rm s})$ does not depend on $x_{\rm c}$ and the phases of $V_g$. Also, it depends on $(\hat{u},\hat{v})$ only through the new phase-space variables $\hat{u}'=\hat{u}+\hat{v}$ and $\hat{v}'=\hat{v}-\hat{u}$. This implies that the unit cell of periodicity of $\hat{H}_1(\hat{u},\hat{v};\hbar_{\rm s})$ in phase space is half the size of the ordinary $2\pi\times 2\pi$ unit cell. Classically, this means that the SWC web cell is expected to be approximately half the size of the ordinary weak-chaos web cell. This is demonstrated by the example in Fig. 1(b) (to be compared with Fig. 1(a)), showing the case of $\ell ' =3>2N=2$ for $N=1$.

Quantally, the eigenvalues of the operator (\ref{Ures}) only approximate the exact eigenvalues $\exp (-i{\cal E})$, so that the scaled QE ${\cal E}/(\mu\epsilon )$ should approximate the eigenvalues of $\hat{H}_1(\hat{u},\hat{v};\hbar_{\rm s})$. The latter operator depends on the variables above which satisfy $[\hat{u}',\hat{v}']=2\pi i\hbar_{\rm s}'$, where $\hbar_{\rm s}'=2\hbar_{\rm s}$. Now, the exact QE spectrum of the operator (\ref{Ure}) at fixed $\mu$ is periodic in $\hbar_{\rm s}$ with period $1$ since the matrices in Eq. (\ref{McUr}) exhibit this periodicity, as one can see from Eq. (\ref{M_elements}). Similarly, the spectrum of $\hat{H}_1(\hat{u},\hat{v};\hbar_{\rm s})$ must be periodic in $\hbar_{\rm s}'$ with period $1$. But when $\hbar_{\rm s}$ covers the interval $0\leq \hbar_{\rm s}<1$, $\hbar_{\rm s}'=2\hbar_{\rm s}$ will cover the interval $0\leq \hbar_{\rm s}'<2$. Therefore, the scaled QE spectrum ${\cal E}/(\mu\epsilon )$ at fixed $\mu$ should exhibit an approximately double structure, i.e., this spectrum  for $0\leq \hbar_{\rm s}<1/2$ should look almost the same as that for $1/2\leq \hbar_{\rm s}<1$.       

Let us show this explicitly in the case of standard potentials (\ref{V}) with $N=1$ and $\ell '>2N=2$; in this case, other results can be derived. Choosing, for the sake of definiteness, $|V_1|=1/2$ [corresponding to the cosine potential $V(x)=\cos (x+\alpha )$ for arbitrary phase $\alpha$], Eq. (\ref{h1_generic}) can be rewritten in this case as: 
\begin{equation}\label{msh1}
\frac{2\cos (\eta )\hat{H}_1(\hat{u},\hat{v};\hbar_{\rm s})}{\ell '} = -\left [\cos \left(\hat{u}'\right) + \cos \left(\hat{v}'\right)\right] ,
\end{equation}
using $J(1;\hbar_{\rm s})=1$ from Eq. (\ref{id1}). Since the eigenvalues of $\hat{H}_1(\hat{u},\hat{v};\hbar_{\rm s})$ are approximately ${\cal E}/(\mu\epsilon )$ (see above), the eigenvalues of the operator on the left-hand side of Eq. (\ref{msh1}) will approximate $\tilde{\cal E}=2\cos (\eta ){\cal E}/(\ell '\mu\epsilon )=8\cos (\eta ){\cal E}/(r\mu\epsilon )$. The operator on the right-hand side of Eq. (\ref{msh1}) is the Harper one \cite{hm,drh}, whose spectra for all $\hbar_{\rm s}'$ in the interval $0\leq \hbar_{\rm s}'<1$ form the well known Hofstadter ``butterfly" \cite{drh}. Therefore, by the above general considerations, the exact spectra $\tilde{\cal E}$ at fixed $\mu$ will be approximated by a \emph{double} Hofstadter butterfly, as shown in Fig. 2 for several values of $\eta$. In terms of the variable $\tilde{\cal E}$, obtained by scaling the QE ${\cal E}$ by quantities including $\cos (\eta )$, the spectra assume an almost universal form, nearly independent of $\eta$. 

Measures of the small deviations of the exact spectra from the universal form of the double Hofstadter butterfly for small $\mu$ are studied in some detail in Appendix C; we briefly summarize here the main results. While the exact spectra for $\hbar_{\rm s}=0,1$ coincide with those from Eq. (\ref{msh1}), there will be small differences between the two spectra at general values of $\hbar_{\rm s}$. In particular, the exact spectra slightly depend on $x_{\rm c}$ [due to high-order terms in the expansion (\ref{Hee})], unlike the spectra from Eq. (\ref{msh1}). For example, for $\hbar_{\rm s}=1/2$ the difference between the width of the exact spectrum and that of the approximate one (which coincides with that for $\hbar_{\rm s}=0,1$) is an expansion in powers of $\mu$ starting from $\mu ^2$ and featuring a dependence on $x_{\rm c}$ from sufficiently high-order terms. Also, the exact spectrum for $\eta\neq 0$ and $\hbar_{\rm s}=1/2$ generally consists of two bands separated by a small gap, which is visible in Fig. 2(a).
\begin{figure}[tbp]
\includegraphics[width=8.2cm]{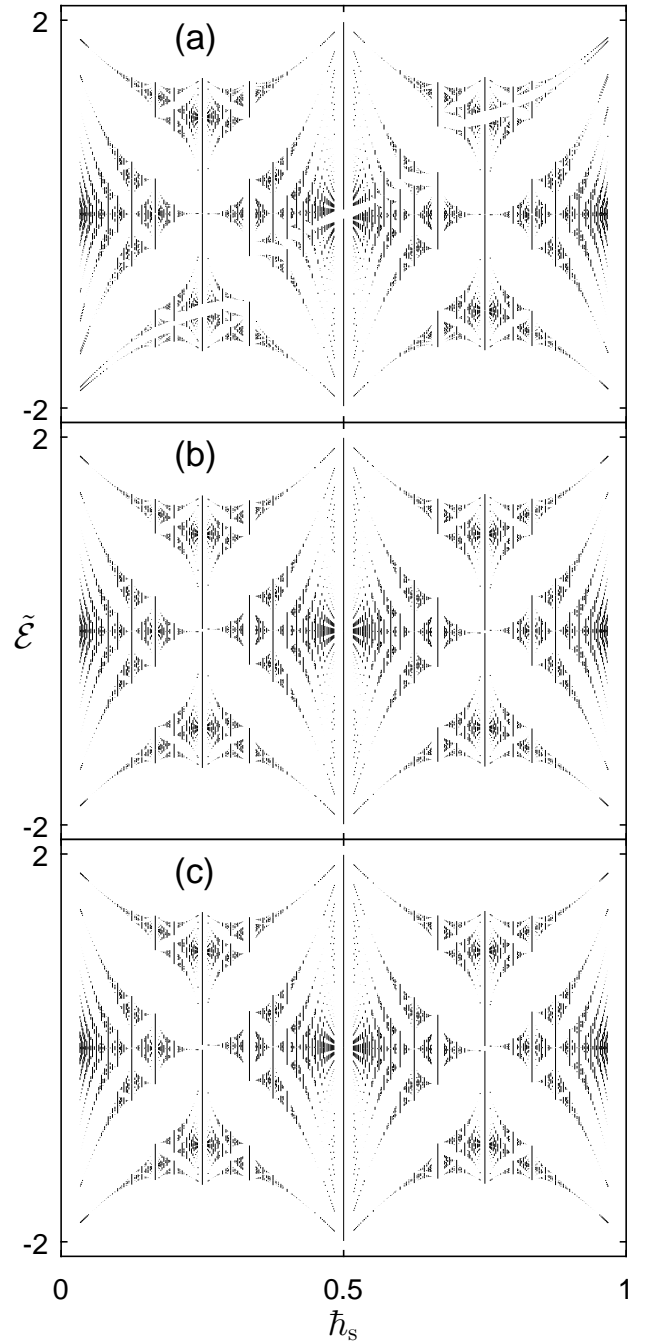}
\caption{(Color online) Scaled QE spectra $\tilde{\cal E}=8\cos (\eta ){\cal E}/(r\mu\epsilon )$ for $\mu =0.1$, $x_{\rm c}=0$, and (a) $\eta /(2\pi )=2/3$; (b) $\eta /(2\pi )=3/5$; (c) $\eta /(2\pi )=8/13$. The scaled Planck constant $\hbar_{\rm s}$ takes the rational values $q/p$ for all coprime integers $p$ and $q$ with $p\leq 30$ and $q<p$. The approximate structure of a double Hofstadter butterfly is evident in all the plots.}
\end{figure}

\begin{center}
\textbf{VI. QUANTUM EVOLUTION AND DIFFUSION}
\end{center}

In this section, we study the time evolution of wave packets and expectation values in semiclassical weak-chaos and SWC regimes for a generic, irrational value of $\hbar_{\rm s}$. Let us apply $s'$ times the basic evolution operator (\ref{Ur}) to an initial wave packet $\Phi_0(u)$ in the $u$ representation. Since the operator (\ref{Ur}) corresponds to $r$ time steps (kicks), the final wave packet will be labeled by $s=rs'$ time steps:
\begin{equation}\label{Prs}
\Phi_{s=rs'}(u) =\hat{U}_r ^{s'}\Phi_0(u) .
\end{equation}

As in Sec. VB, we shall consider a $N=1$ potential (\ref{V}), $V(x)=-\cos (x)$, and we shall assume the strong SWC condition $\ell '>2N=2$. Then, Eq. (\ref{msh1}) holds, so that, using Eqs. (\ref{Ures}), (\ref{msh1}), and $\ell '=r/4$, we get
\begin{equation}\label{Uress}
\hat{U}_r ^{s'}\approx\exp\left\{ i\mu\epsilon\frac{rs'}{8\cos (\eta )}\left[ \cos (\hat{u}')+\cos (\hat{v}')\right]\right\} .
\end{equation}
Defining the scaled time variable
\begin{equation}\label{tau}
\tau =\frac{rs'}{8|\cos (\eta )|}=\frac{s}{8|\cos (\eta )|},
\end{equation}
we see that the approximate evolution operator (\ref{Uress}) exhibits, in terms of $\tau$, a universal form independent of $\eta$. This is analogous to the universal double Hofstadter butterfly in terms of the scaled QE $\tilde{\cal E}=8\cos (\eta ){\cal E}/(r\mu\epsilon )$, see Sec. VB and Fig. 2. Indeed, one may include in $\tau$ a factor $\mu\epsilon$, in analogy to $\tilde{\cal E}$, so that the operator (\ref{Uress}) will be approximately independent also of $\mu$ and $\epsilon$ for small $\mu$. We shall not consider here this generalized definition of $\tau$, since we shall not vary $\mu$ and $\epsilon$.

To verify the universality predicted by Eq. (\ref{Uress}) with Eq. (\ref{tau}) and to study other topics, we calculate expectation values in the evolving state (\ref{Prs}) as functions of $s$ using the exact evolution operator (\ref{Ur}). These calculations can be easily performed by well known methods \cite{fgr,dd1}, which we briefly summarize in Appendix D.   

As the initial wave packet, we choose a normalized coherent state centered at a hyperbolic fixed point $z'=u'+iv'$ of the basic classical map (\ref{cMhb}) (i.e., $z_r=z_0$ for $z_0=z'$): 
\begin{equation}\label{csu}
\Phi_0(u)
=(\pi\hbar)^{-1/4}\exp\left[ iv'u/\hbar-\left(u-u' \right)^2/(2\hbar)\right].  
\end{equation} 
Denoting $\delta\hat{u}=\hat{u}-u'$ and $\delta\hat{v}=\hat{v}-v'$, we consider the expectation value
\begin{equation}\label{evs}
\left\langle \delta\hat{u}^2+\delta\hat{v}^2 \right\rangle_s = \int_{-\infty}^{\infty}du \Phi_{s}^{*}(u)\left( \delta\hat{u}^2+\delta\hat{v}^2 \right)\Phi_{s}(u)
\end{equation}
($\hat{v}=-i\hbar d/du$) in the evolving state (\ref{Prs}) with (\ref{csu}). A classical quantity analogous to (\ref{evs}) is 
\begin{equation}\label{cevs}
\left\langle \left( u_{s}-u_0\right)^2+\left( v_{s}-v_0\right)^2\right\rangle ,
\end{equation}
where $(u_s,v_s)$ ($s=rs'$) is determined from the map (\ref{cMhb}) and $\langle\ \rangle$ denotes average over an ensemble of initial conditions $(u_0,v_0)$ uniformly distributed over a disk centered at $(u',v')$ and of radius $\sqrt{2\hbar}$. Since $(u',v')$ is a hyperbolic fixed point from which there emanates a stochastic web (see Fig. 1), the quantity (\ref{cevs}) will reflect both the classical chaotic diffusion on the web and the stable (elliptic) motions near the web. Our calculations of (\ref{evs}) and (\ref{cevs}) where performed for $\mu =0.1$, $\hbar_{\rm s}=1/[11+(\sqrt{5}-1)/2]$ (corresponding to $\kappa\approx 0.054$), and for several values of $\eta /(2\pi)$ and $x_{\rm c}$. 

The results are shown in Fig. 3. In Fig. 3(a), the quantities (\ref{evs}) and (\ref{cevs}) are plotted in the ordinary weak-chaos case of $\eta/(2\pi )=0/1\ (=0)$ and in the SWC case of $\eta/(2\pi )=2/3$, with $x_{\rm c}=0$ in both cases. The plots of (\ref{evs}) (thick solid line and dashed line) start with a transient behavior almost coinciding with the classical quantity (\ref{cevs}) (thin solid lines) up to some crossover time $s\sim s^{*}$. For $s>s^{*}$, the quantity (\ref{cevs}) saturates to a constant value, due to the fact that for the small value of $\kappa\approx 0.054$ the classical chaotic diffusion is very close to a regular motion on separatrix lines approximating the stochastic web and connecting the hyperbolic fixed points (see Fig. 1). This motion essentially stops when $\delta u$ and $\delta v$ are of the order of the size of the web cell, i.e., $\delta u,\ \delta v\sim 2\pi$ in the weak-chaos case (Fig. 1(a)) and $\delta u,\ \delta v\sim \pi$ in the SWC case (Fig. 1(b)). On the other hand, the quantum wave packet continues to spread because of tunneling between neighboring web cells, leading to quantum diffusion of (\ref{evs}) for $s>s^{*}$. The quantum diffusion is due to the fractal nature of the spectrum  (approximately given by a double Hofstadter butterfly, see Fig. 2) for irrational $\hbar_{\rm s}$ \cite{fqd,wa}. As expected, the quantum-diffusion rate in the weak-chaos case is significantly larger than that in the SWC case. 

Figure 3(b) shows, for $\eta /(2\pi )=2/3$ and $s$ not too large, that the quantities (\ref{evs}) for two extreme values of $x_{\rm c}$ almost coincide, in consistency with the approximate evolution operator (\ref{Uress}), which is independent of $x_{\rm c}$. The dependence on $x_{\rm c}$ emerges only at large $s$, due to high-order terms in the expansion (\ref{Hee}). A similar independence on $x_{\rm c}$ is featured by the corresponding classical quantities (\ref{cevs}) (almost coinciding thin solid lines in Fig. 3(b)), due to an analog of Eq. (\ref{Uress}) for the classical map (\ref{cMhb}) \cite{bhd}.

Figure 3(c) shows the quantities (\ref{evs}) for $\eta /(2\pi )=2/3,3/5,8/13$ and $x_{\rm c}=0$. The closeness of these quantities to the corresponding classical ones (\ref{cevs}), up to some crossover time $s\sim s^{*}$, can now be seen for different SWC values of $\eta$. Again, for $s>s^{*}$, the classical quantities (\ref{cevs}) saturate while the quantum ones (\ref{evs}) feature quantum diffusion. If all these quantities are plotted versus the scaled time variables (\ref{tau}), as in Fig. 3(d), we get an almost perfect coincidence of the quantities (\ref{evs}) for all values of $\eta$, provided $s$ is not too large. This is in agreement with the predictions from Eqs. (\ref{Uress}) and (\ref{tau}); the deviations from coincidence for large $s$ are due to high-order terms in the expansion (\ref{Hee}). The very good coincidence of the classical quantities (\ref{cevs}) for all $\eta$ is again due to the analog of Eq. (\ref{Uress}) for the classical map (\ref{cMhb}).    

\begin{figure}[tbp]
\includegraphics[width=8.2cm]{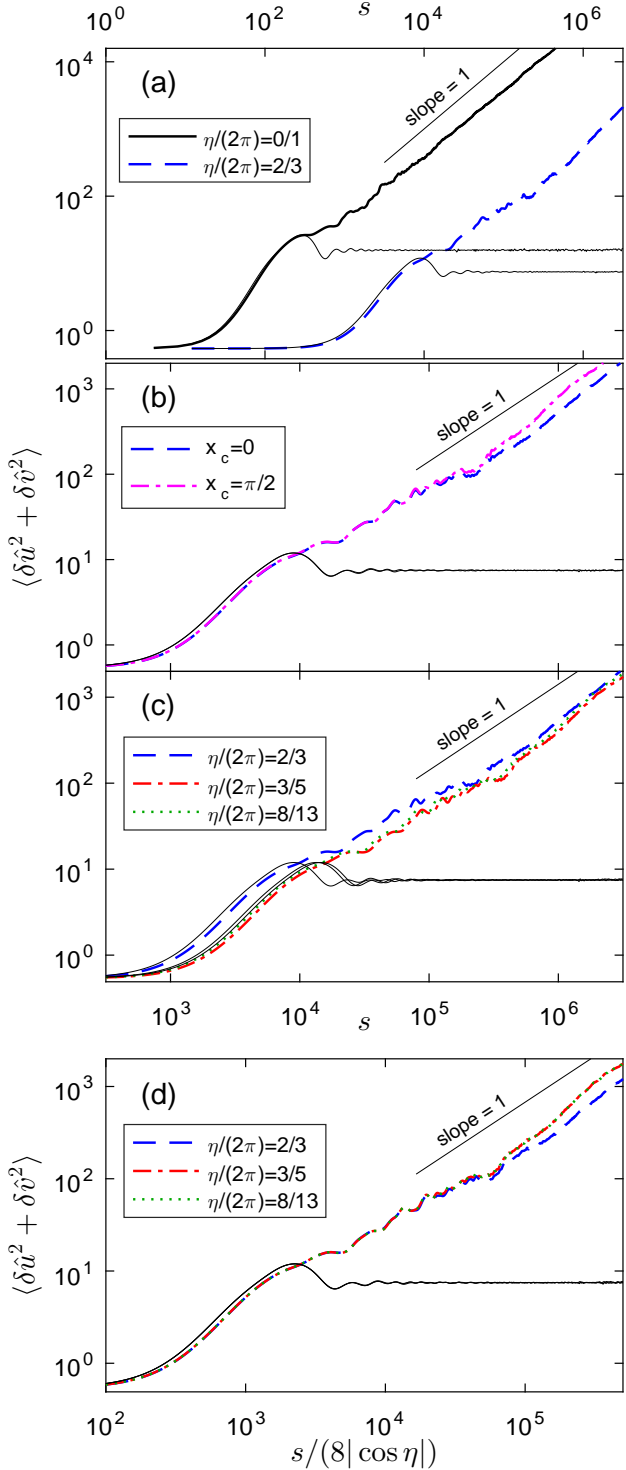}
\caption{(Color online) Plots of the quantities (\ref{evs}) for $\mu =0.1$, $\hbar_{\rm s}=1/[11+(\sqrt{5}-1)/2]$, and several values of $\eta /(2\pi )$ or $x_{\rm c}$, as specified in the legends. The classical quantities (\ref{cevs}) [close to their quantum counterparts (\ref{evs}) for small $s$] are plotted as thin solid lines. (a) Cases of $\eta /(2\pi )=0/1, 2/3$ for $x_{\rm c}=0$. (b) Cases of $\eta /(2\pi )=2/3$ for $x_{\rm c}=0,\pi /2$. (c) Cases of $\eta /(2\pi )=2/3,3/5,8/13$ for $x_{\rm c}=0$; the classical quantities (\ref{cevs}) (thin solid lines) correspond to $\eta /(2\pi )=2/3,8/13,3/5$ in order of descending lines at $s=5000$. (d) Same plots as in (c) but versus the scaled time variable (\ref{tau}); there is an almost perfect coincidence of the quantities (\ref{evs}) for the different values of $\eta /(2\pi )$ and even a better coincidence of the corresponding classical quantities (\ref{cevs}).}
\end{figure} 

\begin{center}
\textbf{VII. SUMMARY AND CONCLUSIONS}
\end{center}

In this paper, we have studied quantum-dynamical and spectral manifestations of classical SWC in KHSs, defined by the general Hamiltonian (\ref{H}). The presence of an electric field $E$, satisfying the resonance conditions (\ref{rge}) with $\eta =ET\neq 0$, causes SWC, defined by Eq. (\ref{SWC}), to be a generic phenomenon in KHSs, occurring for arbitrary potential (\ref{V}) with $N< \ell '$ [Eq. (\ref{cE})] and arbitrary initial value $x_{\rm c}^{(0)}$ in Eq. (\ref{xct}) \cite{bhd}.

We have shown that quantum antiresonance (QAR) or frozen quantum dynamics, with the basic KHS evolution operator (\ref{Ure}) identically equal to a constant phase factor, occurs for integer values of a scaled Planck constant $\hbar_{\rm s}$ and under the same generic classical conditions for SWC. Thus, QAR may be viewed as a quantum analog of SWC. In fact, in the close vicinity of QAR ($\hbar_{\rm s}$ close to an integer), the evolution operator (\ref{Ure}) was shown to describe a regime of semiclassical SWC. The generic QAR in KHSs should be compared with the rare one occurring in other systems \cite{fmi,qar,d,dd1}.

A global spectral manifestation of SWC was shown in the general case of $\ell ' >2N$, a condition stronger than the usual SWC one (\ref{cE}): The plot of a scaled QE spectrum versus $\hbar_{\rm s}$ at fixed small value of $\mu =\kappa /\hbar$ exhibits an approximately doubled structure, i.e., it is approximately periodic in $\hbar_{\rm s}$ with period $1/2$ rather than the ordinary period $1$. This reflects the classical fact that the unit cell of the SWC web cell for $\ell ' >2N$ is approximately half the size of the ordinary weak-chaos web cell, see Fig. 1. In the case of standard (cosine) potentials, with $N=1$ and $\ell '>2$, the plot of a scaled QE spectrum is approximately a double Hofstadter butterfly having universal features; see Sec. VB and Fig. 2. 

This universality is reflected in the quantum evolution of wave packets for standard potentials. This evolution was predicted to exhibit an approximately universal behavior, independent of $\eta$ and $x_{\rm c}^{(0)}$, in terms of the scaled time variable (\ref{tau}). We have verified this prediction for times not too large by numerical studies of the evolution of the kinetic-energy expectation value for a small generic (irrational) value of $\hbar_{\rm s}$; see Sec. VI and Fig. 3. 

The generic asymptotic quantum-diffusive behavior for $\ell '>2$ and irrational $\hbar_{\rm s}$, illustrated by Fig. 3, is replaced by other behaviors for $\ell ' =1,2$. For example, one may get an asymptotic quantum ballistic motion in these cases, as clearly shown in Fig. 4.

We remark that since the general KHS is essentially equivalent to a modulated kicked harmonic oscillator [because of Eqs. (\ref{eKHO}) and (\ref{xct})], the quantized KHS may be experimentally realizable using atom-optics methods with cold atoms or Bose-Einstein condensates, as it was done for the ordinary quantum
kicked harmonic oscillator \cite{dmcw}.
\begin{figure}[tbp]
\includegraphics[width=8.2cm]{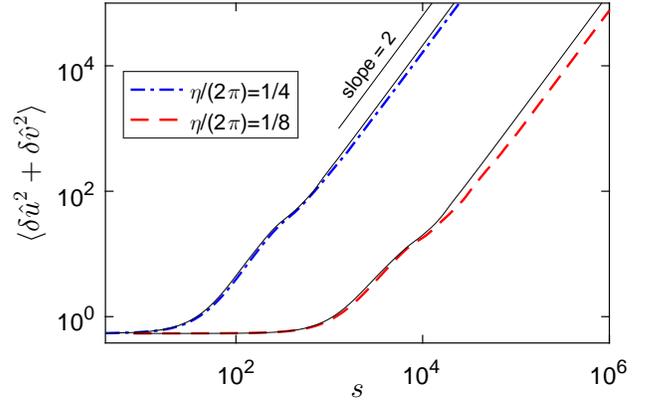}
\caption{(Color online) Same as Fig. 3(a) but in the ordinary weak-chaos case of $\eta /(2\pi )=1/4$ ($\ell ' =1)$ and in the SWC case of $\eta /(2\pi )=1/8$ ($\ell ' =2)$; in both cases, $x_{\rm c}=0.3\pi /2$. It is evident that the asymptotic quantum evolution is now that of a ballistic motion rather than quantum diffusion. Again, however, the rate of this motion in the SWC case is significantly smaller than that in the weak-chaos case. The thin solid lines are the corresponding classical quantities (\ref{cevs}), calculated as explained in Sec. VI, which again agree reasonably well with the quantum behaviors.}
\end{figure}

\renewcommand{\theequation}{A\arabic{equation}}
\setcounter{equation}{0}
\begin{center}
\textbf{APPENDIX A}
\end{center}

We derive here an explicit expression for $\hat{H}_1 (\hat{u},\hat{v};\hbar_{\rm s})$ in Eq. (\ref{Hee}) for the general potential (\ref{V}). From what we mentioned at the end of Sec. IIIB, we see that this expression results entirely from the simple (non-repeated) commutators in Eq. (\ref{id}). Using then Eqs. (\ref{Ur})-(\ref{id}), we can write
\begin{eqnarray}\label{h1_def}
\epsilon \hat{H}_1 (\hat{u},\hat{v};\hbar_{\rm s})
&=&-\frac{1}{2i\mu}
\sum_{j=1}^{r} \sum_{j'=j}^{r} \left[\hat{O}_{j},\hat{O}_{j'}\right] \notag \\ 
&=&-\frac{i\mu}{2}
\sum_{g=-N}^{N} \sum_{g'=-N}^{N} V_{g,g'}(x_{\mathrm{c}})\hat{F}_{g,g'},
\end{eqnarray}
where $V_{g,g'}(x_{\mathrm{c}})=V_g V_{g'} e^{i(g+g')x_{\mathrm{c}}}$ and
\begin{eqnarray}\label{fgg_def}
\hat{F}_{g,g'}=\sum_{j=1}^{r} \sum_{j'=j}^{r} \hat{C}_{g,g'}^{(j,j')}
\end{eqnarray}
with
\begin{eqnarray}\label{c_def}
\notag
\hat{C}_{g,g'}^{(j,j')}
=\left[e^{-ig(j\eta+\hat{v}_j)}, e^{-ig'(j'\eta+\hat{v}_{j'})} \right].
\end{eqnarray}
We decompose $\hat{F}_{g,g'}$ in Eq. (\ref{fgg_def}) into three summations as follows
\begin{eqnarray}\label{fgg_in_three_sum}
\notag
\hat{F}_{g,g'}
&=& \sum_{j=-3}^{r-4} \sum_{j'=j+4}^{r} \hat{C}_{g,g'}^{(j+4,j')} \\
\notag
&=&e^{-i4g\eta}
\sum_{j'=1}^{r} \sum_{j=-3}^{j'-4} \hat{C}_{g,g'}^{(j,j')}\\
\notag
&=&e^{-i4g\eta} \sum_{j'=1}^{r}\left(
 \sum_{j=1}^{j'}+\sum_{j=-3}^{0}-\sum_{j=j'-3}^{j'} \right)
\hat{C}_{g,g'}^{(j,j')} \\
\notag
&=&e^{-i4g\eta}\left(
\sum_{j=1}^{r} \sum_{j'=j}^{r} \hat{C}_{g,g'}^{(j,j')}
+\sum_{j=-3}^{0} \sum_{j'=1}^{r} \hat{C}_{g,g'}^{(j,j')} \right.\\
&-&\left. \sum_{j'=1}^{r} \sum_{j=j'-3}^{j'} \hat{C}_{g,g'}^{(j,j')}
\right), 
\end{eqnarray}
where we used the periodicity of $\hat{v}_j$ in $j$ with period 4 to get the second equality. The first summation in the parentheses of Eq. (\ref{fgg_in_three_sum}) is $\hat{F}_{g,g'}$ itself [see Eq. (\ref{fgg_def})] and the second one does not contribute since $\sum_{j'=1}^{r} e^{-ig'(j'\eta+\hat{v}_{j'})}=0$ under the SWC condition $\ell ' > N$, as shown in the derivation of Eq. (\ref{H0e}). Therefore, $\hat{F}_{g,g'}$ is expressed by the third summation as  
\begin{eqnarray}\label{fgg_in_3rd_one}
\notag
\hat{F}_{g,g'}
&=&\frac{e^{-i4g\eta}}{e^{-i4g\eta}-1}
\sum_{j'=1}^{r} \sum_{j=j'-3}^{j'} \hat{C}_{g,g'}^{(j,j')} \\ \notag
&=&\frac{1}{e^{-i4g\eta}-1}
\sum_{l=0}^{\ell ' - 1}e^{-i4(g+g')\eta l} 
\sum_{\bar{n}=1}^4 e^{-ig\bar{n}\eta} \\
&\times& \sum_{\bar{n}'=1}^4
e^{-i(g+g')\bar{n}'\eta}
\left[e^{-ig\hat{v}_{\bar{n}+\bar{n}'}}, e^{-ig'\hat{v}_{\bar{n}'}} \right], 
\end{eqnarray}
where the second equality is obtained after the successive replacements 
$j \to j'+\bar{n} -4$, $j' \to 4l +\bar{n}'$, and by using the periodicity of $\hat{v}_j$ in $j$ with period 4 in $\hat{C}_{g,g'}^{(j,j')}$. For $\ell ' > N$, one has $|g+g'|<2\ell '$, so that the geometric sum over $l$ in Eq. (\ref{fgg_in_3rd_one}) gives  
\begin{eqnarray}\label{three-delta}
\sum_{l=0}^{\ell ' -1}e^{-i4(g+g')\eta l} 
= \ell ' \left( \delta_{g+g',0} + \delta_{g+g',\ell '} 
+ \delta_{g+g',-\ell '}\right).
\end{eqnarray}
Using Eqs. (\ref{three-delta}) and (\ref{c1}), 
and also the fact that the commutator in Eq. (\ref{fgg_in_3rd_one}) does not 
vanish only for $\bar{n} =1,3$, we find, after some lengthy but straightforward calculations, 
\begin{eqnarray}\label{fgg_ff}
\hat{F}_{g,g'}=\frac{1}{2}\left( \hat{f}_{g,g'} + \hat{f}_{g',g} \right), 
\end{eqnarray}
where 
\begin{eqnarray}\label{fgg}
\notag
\hat{f}_{g,g'} 
&=&-4\ell ' \frac{\sin (gg'\pi \hbar_{\rm s})}{\sin(2g\eta)}  
\left( \delta_{g+g',0} + \delta_{g+g',\ell '} + \delta_{g+g',-\ell '}\right) \\ 
\notag
& \times & \Bigl[ e^{-ig\eta} 
\cos \bigl( g'\hat{u}+g\hat{v} +(g+g')\eta \bigr)\Bigr. \\ 
& - & \Bigl. e^{ig\eta}\cos \bigr( g' \hat{u}-g\hat{v}+(g+g')\eta \bigr) \Bigr].
\end{eqnarray}
Note that the denominator $\sin(2g\eta)$ in Eq. (\ref{fgg}) does not vanish, 
since its argument $2g\eta = \pi k n' g/\ell ' $ is never an integer multiple 
of $\pi$ except for $g=0$, which should be neglected since we take $V_0=0$ 
by definition in Eq. (\ref{V}). Substituting Eq. (\ref{fgg_ff}) with (\ref{fgg}) into (\ref{h1_def}) and using (\ref{id1}), we obtain the explicit expression for $\hat{H}_1(\hat{u},\hat{v};\hbar_{\rm s})$:
 \begin{eqnarray}\label{h1_generic}
\notag
& & \hat{H}_1(\hat{u},\hat{v};\hbar_{\rm s}) = -2\ell '\sum_{g=1}^N J(g^2;\hbar_{\rm s})
\frac{|V_g|^2}{\cos (g\eta)} \\ \notag
& \times & \bigl\lbrace\cos\left[g\left(\hat{u}+\hat{v}\right)\right] + 
\cos\left[g\left(\hat{u}-\hat{v}\right)\right]\bigr\rbrace  \\ \notag
& - & 4\ell '\Im \Biggl \lbrace  e^{i\ell 'x_{\mathrm{c}}}\sum_{g=\ell '-N}^N 
J\left[g(\ell '-g);\hbar_{\rm s}\right]\frac{V_gV_{\ell '-g}}{\sin (2g\eta )} \Biggr. \\ \notag
& \times & \Bigl[ e^{-ig\eta}\cos \bigl( \left(\ell '-g \right)\hat{u}+g\hat{v} +\ell '\eta \bigr)\Bigr. \\  
& - & \Biggl. \Bigl. e^{ig\eta}\cos \bigl( \left( \ell '-g\right)\hat{u}-g\hat{v}+\ell '\eta \bigr) \Bigr] \Biggr \rbrace.
\end{eqnarray}
We see from Eq. (\ref{h1_generic}) that the dependence on $x_{\mathrm{c}}$  
in $\hat{H}_1(\hat{u},\hat{v};\hbar_{\rm s})$ arises only from the second sum, which does not appear for $\ell ' >2N$.

\renewcommand{\theequation}{B\arabic{equation}}
\setcounter{equation}{0}
\begin{center}
\textbf{APPENDIX B}
\end{center}

We derive here the matrix elements of ${\mathbf M}_r({\mathbf w})$ in Eq. (\ref{McUr}). To this end, we start from a generalized kicked-Harper evolution operator  \cite{d1} 
defined by 
\begin{eqnarray}\label{UGen}
\hat{U}_{\rm KH}
=\exp\left[{-i\mu W_2(\hat{v})}\right]\exp\left[{-i\mu W_1(\hat{u})}\right], 
\end{eqnarray}
where $W_1(x)$ and $W_2(x)$ are arbitrary $2\pi$-periodic functions.  
Each factor in $\hat{U}_{\rm KH}$ can be expanded in a Fourier series as 
\begin{eqnarray}\label{W_fourier}
\notag
\exp\left[{-i\mu W_1(\hat{u})}\right]&=&\sum_{s=-\infty}^{\infty}J_{1,s}e^{is\hat{u}},\\
\exp\left[{-i\mu W_2(\hat{v})}\right]&=&\sum_{s=-\infty}^{\infty}J_{2,s}e^{is\hat{v}}.  
\end{eqnarray}
By applying the operator (\ref{UGen}) to the states (\ref{qes}) and using Eq. (\ref{W_fourier}), we get
\begin{eqnarray}\label{UgenPsi}
\notag
&&\langle v|\hat{U}_{\rm KH}|\Psi_{b,\mathbf{w}}\rangle
=\sum_{l,l'=0}^{p-1}\tilde{F}_{1,l}(w_1)\tilde{F}_{2,l}(w_2)e^{i(l w_1+l' w_2)}\\
\notag
&&~~~~~~~~~~~~\times \sum_{d=0}^{p-1} \phi_b(d;\mathbf{w})e^{2\pi i l d \hbar_{\rm s}}
\psi_{w_1+2\pi(d-l')\hbar_{\rm s},w_2}(v), \\
\end{eqnarray}
where, for $j=1,2$, 
\begin{eqnarray}\label{F_fn}
\tilde{F}_{j,l}(w_j)=\sum_{s=-\infty}^{\infty}J_{j,sp+l}e^{i s p w_j}, 
\end{eqnarray}
and 
\begin{eqnarray}\label{kq_fn}
\psi_{\mathbf{w}}(v)=\sum_{l=-\infty}^{\infty}e^{i l w_1/q}\delta(v-w_2+2\pi l/p)  
\end{eqnarray}
are ``$kq$" distributions \cite{jz}. Defining 
\begin{eqnarray}\label{phi_bar}
\bar{\phi}_b(d;\mathbf{w}) \equiv e^{i d w_2}\phi_b(d;\mathbf{w}), 
\end{eqnarray}
and using the independence of the $kq$ distributions (\ref{kq_fn}) for different $\mathbf{w}$'s \cite{jz}, we obtain from Eq. (\ref{UgenPsi}) the eigenvalue equation for the 
vector $\mathbf{V}_b(\mathbf{w})\equiv \{\bar{\phi}_b(d;\mathbf{w})\}_{d=0}^{p-1}$:
\begin{eqnarray}\label{MV}
\mathbf{M}_{\rm KH}(\mathbf{w})\mathbf{V}_b(\mathbf{w})
=\exp[-i{\cal E}_b(\mathbf{w})]\mathbf{V}_b(\mathbf{w}), 
\end{eqnarray}
where $\mathbf{M}_{\rm KH}(\mathbf{w})$ is a $p\times p$ unitary matrix with elements
\begin{eqnarray}\label{M_elements}
\notag
(\mathbf{M}_{\rm KH})_{d,d'}(\mathbf{w})
&=&\frac{1}{p}\sum_{s=0}^{p-1}\exp\Big[-i\mu W_1(w_1+2\pi\hbar_{\rm s}d') \Big. \\
&-& i\mu W_2 (w_2+  2\pi s/p ) \\
\notag
\Big.&-& i(w_2+ 2\pi s/p)(d'-d)\Big], 
\end{eqnarray}
$d, d'=0,...,p-1$. The matrix ${\mathbf M}_r({\mathbf w})$ in Eq. (\ref{McUr}) is the product of matrices ${\mathbf M}^{(j)}_{\rm KH}({\mathbf w})$ having elements (\ref{M_elements}) with 
\begin{eqnarray}\label{W1_and_W2}
\notag
W_1(x) &=& V\big(x_{\rm c}-2(j+1)\eta-(-1)^j x\big) , \\
W_2(x) &=& V\big(x_{\rm c}-2j\eta-(-1)^j x\big).
\end{eqnarray}
Since the eigenvalue equation for each matrix ${\mathbf M}^{(j)}_{\rm KH}({\mathbf w})$ has the form (\ref{MV}), also the eigenvalue equation for the matrix (\ref{McUr}) will have this form, where $\mathbf{V}_b(\mathbf{w})\equiv \{\bar{\phi}_b(d;\mathbf{w})\}_{d=0}^{p-1}$ and $\bar{\phi}_b(d;\mathbf{w})$ are given by Eq. (\ref{phi_bar}) with $\phi_b(d;\mathbf{w})$ being the coefficients giving the eigenstates (\ref{qes}).  

\renewcommand{\theequation}{C\arabic{equation}}
\setcounter{equation}{0}
\begin{center}
\textbf{APPENDIX C}
\end{center}

We evaluate here the spectrum width and gap width for $\hbar_{\rm s}=1/2$ 
and several SWC values of $\eta$ in the case of the potential $V(x)=-\cos(x)$. 
For $\hbar_{\rm s}=1/2$ ($p=2$), ${\mathbf M}_{r}({\mathbf w})$ 
in Eq. (\ref{McUr}) is the product of $2 \times 2$ matrices  
${\mathbf M}^{(j)}_{\rm KH}({\mathbf w})$ having elements 
(\ref{M_elements}) with (\ref{W1_and_W2}). For $V(x)=-\cos(x)$, these matrices can be compactly written as    
\begin{eqnarray}\label{MjKH}
\notag
{\mathbf M}^{(j)}_{\rm KH}({\mathbf w})
&=&{\mathbf D}_{w_2}e^{i\mu \cos\left[x_{\mathrm{c}}-2j\eta - (-1)^j w_2\right]\sigma_x}\\
&\times& e^{i\mu \cos\left[x_{\mathrm{c}}-(2j+1)\eta - (-1)^j w_1 \right]\sigma_z}
{\mathbf D}_{w_2}^{-1}, 
\end{eqnarray}
where $\sigma_x$ and $\sigma_z$ are Pauli matrices and ${\mathbf D}_{w_2}={\rm diag}(1,e^{i w_2})$. Equations (\ref{McUr}) and (\ref{MjKH}) imply that $\det {\bf M}_r({\bf w}) = 1$. Thus, if the QE eigenvalues of ${\mathbf M}_{r}({\mathbf w})$ are $\exp[{-i{\cal E}_1({\bf w})}]$ and $\exp[{-i{\cal E}_2({\bf w})}]$, we must have ${\cal E}_1({\bf w})=-{\cal E}_2({\bf w})$. Also, $\mathrm{Tr}\left[{\bf M}_r({\bf w})\right]=2\cos[\Delta {\cal E}({\bf w}) /2]$, where $\Delta {\cal E}({\bf w})={\cal E}_1({\bf w})-{\cal E}_2({\bf w})=2{\cal E}_1({\bf w})$.  

For $\eta=0$ ($r=4$, $\ell '=1$), SWC occurs only at $x_{\rm c} =\pi/2$. In this case, we get from Eqs. (\ref{McUr}) and (\ref{MjKH}) the exact result   
\begin{eqnarray}\label{qe_diff_eta0}
\Delta {\cal E}({\bf w}) = 4\arcsin\{\sin[\mu\sin (w_1)]
\sin[\mu\sin(w_2)]\}. 
\end{eqnarray}
It is clear from Eq. (\ref{qe_diff_eta0}) that the spectrum width $\Delta =\Delta {\cal E}(\pi /2,\pi /2)$ and the gap width $\delta =\Delta {\cal E}({\bf w}')$, where ${\bf w}'=(w_1,0),(0,w_2)$:   
\begin{eqnarray}\label{width_gap_01}
\Delta
&=& 4\arcsin \left[\sin^2(\mu)\right]
= 4\mu ^2(1-\mu^2/3+\cdots ), \notag \\
\delta &=& 0.
\end{eqnarray}
Thus, there is no gap in this case. 

For $\eta \neq 0$ with odd $\ell '$, $\mathrm{Tr}\left[{\bf M}_r({\bf w})\right]$ has symmetries in the Brillouin zone of ${\bf w}$. This can be shown by the following calculation, using Eq. (\ref{McUr}) with $r=4\ell '$:    
\begin{eqnarray}\label{sym_deriv}
\mathrm{Tr}\left[{\bf M}_r(-{\bf w})\right]
&=& \mathrm{Tr}\left[\prod_{j=0}^{2\ell '-1} {\bf M}^{(j)}_{\rm KH}(-{\bf w})\right]\\
\notag
&=& \mathrm{Tr}\left[ 
\prod_{j=0}^{\ell '-1} {\bf M}^{(j)}_{\rm KH}(-{\bf w})
\prod_{j=\ell '}^{2\ell '-1} {\bf M}^{(j)}_{\rm KH}(-{\bf w})\right]\\
\notag
&=& \mathrm{Tr}\left[ 
\prod_{j=\ell '}^{2\ell '-1} {\bf M}^{(j-\ell ')}_{\rm KH}(-{\bf w})\prod_{j=0}^{\ell '-1} {\bf M}^{(j+\ell ')}_{\rm KH}(-{\bf w})
\right].
\end{eqnarray}
In the last line of Eq. (\ref{sym_deriv}), the shift by $\pm \ell '$ in $j$ gives both a sign change ${-\bf w}\to{\bf w}$ and a phase shift by  
$2 \ell ' \eta =  \pi n' k$ in the cosine functions in Eq. (\ref{MjKH}). 
For even $n' k$, this shift can be ignored; we then get, using also the identity $\mathrm{Tr}({\bf A}{\bf B})=\mathrm{Tr}({\bf B}{\bf A})$,
\begin{eqnarray}\label{M_mat_sym1}
\mathrm{Tr}\left[{\bf M}_r (-{\bf w})\right]
= \mathrm{Tr}\left[{\bf M}_r ({\bf w})\right]. 
\end{eqnarray}
In the case of odd $n' k$, the phase shift leads to changes of the sign 
of the exponents in Eq. (\ref{MjKH}); however, using the formula 
$e^{-i a \sigma_{x,z}}=\sigma_y e^{i a \sigma_{x,z}}\sigma_y$, for 
any number $a$, and the fact  that $\sigma_y^2$ is the $2\times 2$ identity matrix, we see that Eq. (\ref{M_mat_sym1}) holds also this case. 

For $\hbar_{\rm s}=1/2$, the matrix (\ref{McUr} is periodic in both $w_1$ and $w_2$ with period $\pi$ (see Sec. VA). Therefore, Eq. (\ref{M_mat_sym1}) can be generalized:    
\begin{eqnarray}\label{M_mat_sym2}
\mathrm{Tr}\left[{\bf M}_r({\bf w}_{\rm t}-{\bf w})\right]
= \mathrm{Tr}\left[{\bf M}_r({\bf w})\right],  
\end{eqnarray}
where ${\bf w}_{\rm t}=(0,0),(\pi,0),(0,\pi),(\pi.\pi)$. It follows from Eq. (\ref{M_mat_sym2}) and $\mathrm{Tr}\left[{\bf M}_r({\bf w})\right]=2\cos[\Delta {\cal E}({\bf w})/2]$ that $\Delta {\cal E}({\bf w})$ is symmetric under inversion around four symmetry centers: ${\bf w}_{\rm sc}=(0,0),(\pi /2,0),(0,\pi /2),(\pi /2,\pi /2)$. Each of these symmetry centers should be an extremum point of $\Delta {\cal E}({\bf w})$. In fact, our numerical observations for all the values of $\eta$ considered below indicate that $\Delta {\cal E}({\bf w})$ has a global minimum at ${\bf w}_{\rm sc}=(0,0)$, a global maximum at ${\bf w}_{\rm sc}=(\pi /2,\pi /2)$, and saddles at ${\bf w}_{\rm sc}=(\pi /2,0),(0,\pi /2)$. Therefore, the spectrum width $\Delta =\Delta {\cal E}(\pi /2,\pi /2$) and the gap width $\delta =\Delta {\cal E}(0,0)$. Expressions for the latter quantities were calculated using Mathematica. In terms of the scaled QE $\tilde{\cal E}=2\cos (\eta ){\cal E}/(\ell '\mu\epsilon )$ (see Sec. VB), where $\epsilon =\mu$ for $\hbar_{\rm s}=1/2$, these expressions are given by     
\begin{eqnarray}\label{width_gap_23}
\notag
\tilde{\Delta}
&=& 4\left[ 1 - \mu^2 +\frac{1}{360}\bigl( 685- \cos(6 x_{\mathrm c})\bigr)\mu^4+\cdots \right], \\
\tilde{\delta}
&=& 2\frac{\sqrt{2}}{3}\bigl|\cos(3 x_{\mathrm c})\bigr|\mu 
\left(1- \frac{1}{4}\mu^2 +\cdots \right)
\end{eqnarray}
for $\eta/(2\pi)=2/3$,   
\begin{eqnarray}\label{width_gap_35}
\notag
\tilde{\Delta}
&=& 4\left(1 - \frac{3-\sqrt{5}}{2}\mu^2 +\frac{246-107\sqrt{5}}{36}\mu^4+\cdots \right), \\
\notag
\tilde{\delta}
&=& \frac{\sqrt{2}\left(6+\sqrt{5}\right)}{30}
\bigl|\cos(5 x_{\mathrm c})\bigr|\mu^3\\
&~&~~~~\times \left(1 - \frac{81+2\sqrt{5}}{186} \mu^2 +\cdots \right)
\end{eqnarray}
for $\eta/(2\pi)=3/5$, and 
\begin{eqnarray}\label{width_gap_813}
\notag
\tilde{\Delta}
&=& 4\left(1 - a_2 \mu^2 +a_4 \mu^4+\cdots \right), \\
\tilde{\delta}
&=& 0.00389344\bigl|\cos(13 x_{\mathrm c})\bigr|\mu^{11}+\cdots  
\end{eqnarray}
for $\eta/(2\pi)=8/13$, where $a_2=-0.446215$ and $a_4=0.324429$ are roots of 
6th-degree algebraic equations.  

\renewcommand{\theequation}{D\arabic{equation}}
\setcounter{equation}{0}
\begin{center}
\textbf{APPENDIX D}
\end{center}

We summarize here briefly well known methods \cite{fgr,dd1} for calculating the evolving state (\ref{Prs}) and expectation values in it using the exact evolution operator (\ref{Ur}). Let us first express $\Phi_0(u)$ (assumed to be normalized) by its $v$-representation $\bar{\Phi}_0(v)$:
\begin{eqnarray}\label{DPb}
\Phi_0(u) & = &\hbar ^{-1}\int_{-\infty}^{\infty}dv\exp (iuv/\hbar )\bar{\Phi}_0(v) \notag \\ & = & \int_{0}^{1}d\beta \exp (i\beta u)\Phi_{0,\beta}(u) ,
\end{eqnarray}
\begin{equation}\label{Pb}
\Phi_{0,\beta}(u) =\sum_{l=-\infty}^{\infty} \bar{\Phi}_0[(l+\beta)\hbar]\exp (ilu) .
\end{equation}
Since the function (\ref{Pb}) is clearly $2\pi$-periodic in $u$, Eq. (\ref{DPb}) is a decomposition of $\Phi_0(u)$ into Bloch functions $\exp (i\beta u)\Phi_{0,\beta}(u)$ with quasimomenta $\beta\hbar$, $0\leq\beta<1$. By applying to such a function the evolution operator (\ref{Ur}), denoted here by $\hat{U}_r(\hat{u},\hat{v})$, and using $\hat{v}=-i\hbar d/du$, we easily get:
\begin{equation}\label{UrPb}
\hat{U}_r(\hat{u},\hat{v})e^{i\beta u}\Phi_{0,\beta}(u)=e^{i\beta u}\hat{U}_{r,\beta}(\hat{u},\hat{v})\Phi_{0,\beta}(u) ,
\end{equation}
where
\begin{equation}\label{Ub}
\hat{U}_{r,\beta}(\hat{u},\hat{v})=\hat{U}_r(\hat{u},\hat{v}+\beta\hbar)=
\hat{U}_r(\hat{u},\beta\hbar-i\hbar d/du).
\end{equation}
Then, by applying $s'$ times the operator (\ref{Ur}) to the initial wave packet (\ref{DPb}) and using Eq. (\ref{UrPb}), we obtain
\begin{equation}\label{DPsb}
\Phi_s(u)=\hat{U}_r^{s'}(\hat{u},\hat{v})\Phi_0(u)=\int_{0}^{1}d\beta \exp (i\beta u)\Phi_{s,\beta}(u),
\end{equation}
where $s=rs'$ and 
\begin{eqnarray}\label{Psb}
\Phi_{s,\beta}(u) &=& \hat{U}_{r,\beta}^{s'}(\hat{u},\hat{v})\Phi_{0,\beta}(u) \notag \\
&=& \sum_{l=-\infty}^{\infty} \bar{\Phi}_s[(l+\beta)\hbar]\exp (ilu) .
\end{eqnarray} 
We thus see that the time evolution (\ref{DPsb}) can be decomposed or ``fibrated" \cite{fgr} into independent evolutions (\ref{Psb}) under the operator (\ref{Ub}) at fixed $\beta$. The latter evolutions are relatively easy to calculate, as explained below, since they involve a Fourier series rather than a Fourier transform.  Also, the expectation value of any Hermitian operator function of $\hat{v}$, $F(\hat{v})$, in the evolving state (\ref{DPsb}) can be written as
\begin{equation}\label{evFv}
\left\langle F(\hat{v}) \right\rangle_s = \int_{-\infty}^{\infty}du \Phi_s^{*}(u)F(\hat{v})\Phi_s(u) = \int_0^1 d\beta \left\langle F(\hat{v}) \right\rangle_{s,\beta},
\end{equation}
where
\begin{equation}\label{evFvb}
\left\langle F(\hat{v}) \right\rangle_{s,\beta}=2\pi \sum_{l=-\infty}^{\infty} \left|\bar{\Phi}_s[(l+\beta)\hbar]\right|^2 F[(l+\beta)\hbar],
\end{equation}
namely, the expectation value (\ref{evFv}) can be fibrated into the expectation values (\ref{evFvb}) at fixed $\beta$, $0\leq\beta <1$.

In Eq. (\ref{Psb}), the discrete $v$-representation $\bar{\Phi}_{s}[(l+\beta )\hbar ]$ of the evolving wave packet for $s=rs'+j'$, $j'=0,...,r-1$, can be calculated, for even $j'$, by multiplying $\bar{\Phi}_{0}[(l+\beta )\hbar ]$ by terms $j=r-j'$ in the product (\ref{Ur}) with $\hat{v}$ replaced by $(l+\beta )\hbar$; the application of the $\hat{u}$-dependent terms for odd $j'$ in (\ref{Ur}) is equivalent to a convolution in the discrete $v=(l+\beta)\hbar$ space. We thus have
\begin{eqnarray}\label{bPse}
\bar{\Phi}_{rs'+j'+1}[(l+\beta )\hbar ]
& = & \exp \{ -i\mu V[x_{\rm c}+j'\eta-(-1)^{j'/2} \notag \\
& \times &(l+\beta)\hbar] \}\bar{\Phi}_{rs'+j'}[(l+\beta )\hbar ],
\end{eqnarray}
for $j'=0,2,4,\dots,r-2$ and
\begin{eqnarray}\label{bPso}
\bar{\Phi}_{rs'+j'+1}[(l+\beta )\hbar ]
& = & \sum_{l'=-\infty}^{\infty}
\tilde{J}_{(-1)^{(j'-1)/2}(l-l')}\left(x_{\rm c}+j'\eta;\mu\right) \notag \\
& \times & \bar{\Phi}_{rs'+j'}[(l'+\beta )\hbar ],
\end{eqnarray}
for $j'=1,3,5,\dots,r-1$, where $\tilde{J}_l(x;\mu)$ is defined by
\begin{equation}\label{tJ}
\exp\left[-i\mu V(x+u)\right]=\sum_{l=-\infty}^{\infty}
\tilde{J}_l(x;\mu)\exp(ilu).
\end{equation}
As Fourier coefficients in Eq. (\ref{tJ}), $\tilde{J}_l(x;\mu)$ usually decay fast with $|l|$, so that the sum in Eq. (\ref{bPso}) can be truncated to get accurate enough results in a simple way.

Similarly, the expectation value of any Hermitian operator function of $\hat{u}$, $G(\hat{u})$, in the evolving state can be calculated by fibrating it  into the expectation values at fixed quasiposition $\lambda\hbar$ for all  $\lambda$, $0\leq\lambda <1$; i.e., in Eqs. (\ref{evFv}) and (\ref{evFvb}), one essentially replaces $\hat{v}$, $u$, $\beta$ by $\hat{u}$, $v$, $\lambda$, respectively, exchanging also $\Phi_s$ and $\bar{\Phi}_s$. One can then calculate expectation values such as (\ref{evs}) in Sec. VI.

\end{document}